\documentclass[onecolumn]{revtex4}

\usepackage{graphicx}% Include figure files
\usepackage{dcolumn}% Align table columns on decimal point
\usepackage{color}
\usepackage{bm}% bold math
\usepackage{amsmath}
\usepackage{amsfonts}
\usepackage{amssymb}

\input epsf

\begin{document}

\title{Universal thermodynamics in different gravity theories: Modified entropy on the horizons}

\author{Saugata Mitra\footnote{saugatamitra20@gmail.com}}
\affiliation{Department of Mathematics, Jadavpur University,\\
 Kolkata-700032, West Bengal, India.}
\author{Subhajit Saha\footnote {subhajit1729@gmail.com}}
\affiliation{Department of Mathematics, Jadavpur University,\\
 Kolkata-700032, West Bengal, India.}
\author{Subenoy Chakraborty\footnote {schakraborty.math@gmail.com}}
\affiliation{Department of Mathematics, Jadavpur University,\\
 Kolkata-700032, West Bengal, India.}

%%%%%%%%%%%%%%%%%%%%%%%%%%%%%%%%%%%%%%%%%%%%%%%%%%%%%%%%%%%%%%%%%%%%%%%%%%%%%%%%%%%%%%%%%%%%%%%%%%%%%%%%%%%%%%%%%%%%%%%%%%%%
\begin{abstract}

The paper deals with universal thermodynamics for FRW model of the universe bounded by apparent (or event) horizon.
Assuming Hawking temperature on the horizon, the unified first law is examined on the horizon for different gravity theories.
The results show that equilibrium configuration is preserved with a modification to Bekenstein entropy on the horizon.

\end{abstract}

\maketitle
%%%%%%%%%%%%%%%%%%%%%%%%%%%%%%%%%%%%%%%%%%%%%%%%%%%%%%%%%%%%%%%%%%%%%%%%%%%%%%%%%%%%%%%%%%%%%%%%%%%%%%%%%%%%%%%%%%%%%%%%%%%%
%\pacs{04.70.Dy, 04.90+e}
%%%%%%%%%%%%%%%%%%%%%%%%%%%%%%%%%%%%%%%%%%%%%%%%%%%%%%%%%%%%%%%%%%%%%%%%%%%%%%%%%%%%%%%%%%%%%%%%%%%%%%%%%%%%%%%%%%%%%%%%%%%%

It is well known today that recent observational predictions \cite{Riess} divide the physicists into two groups.
The first group has been trying to explain this late time acceleration within the frame work of standard cosmology,
assuming the existence of an exotic matter with negative pressure (called dark energy (DE)).
But till now the nature of DE is completely unknown to us and is an unresolved problem in modern
theoretical physics (see \cite{Padmanabhan03}, \cite{Sahni} {\textrm{and references therein}). On the otherhand, the second group is of
the opinion of a modified gravity theory- a modification of Einstein's general relativity. A
common and widely used modified theory is $f(R)$-gravity theory where the Lagrangian density
$R$ (the Ricci scalar) in the Einstein-Hilbert action is replaced by an arbitrary function of
$R$ {\it i.e.,} $f(R)$ (\cite{Sotiriou} for a review and references therein). Also there are other
modified gravity theories namely Scalar Tensor Theory, Brane world scenario, $f(G)$, $f(R,G)$
and $f(T)$ gravity theories, where T is the usual torsion scalar,
$G=R_{ \mu \gamma \rho \sigma } R^{ \mu \gamma \rho \sigma } -4 R_{ \mu \gamma } R^{ \mu \gamma } +R^2 $
is the Gauss Bonnet invariant term and $R_{ \mu \gamma \rho \sigma }$ and $R_{ \mu \gamma }$
are the usual Riemann curvature tensor and Ricci tensor respectively.
These modified theories [5-10] are considered as gravitational alternatives for DE and may serve as dark matter \cite{Sobouti}.

Further inspection of a gravity theory from thermodynamical view point is also an interesting issue in modern theoretical physics.
 The deep connection between gravity and thermodynamics is strongly believed due to Ads/CFT correspondence [12] and black hole thermodynamics [13].
 This belief was put one step forward by the seminal works of Jacobson [14] and Padmanabhan [15]. By introducing the local Rindler horizon and
 assuming the Clausius relation  $\delta Q = TdS$ for all local Rindler causal horizons through each spacetime point, Jacobson deduced the
Einstein's field equations from the proportionality of entropy ($S$) to the horizon area ($A$). Here $ \delta Q $ stands for the variation of
the heat flow and $T$ is the Unruh temperature  measured by an accelerated observer just inside the horizon.
 Although Jacobson derived the equivalence along null directions but it is speculated that the results may also be true along any other direction
 in the tangent to the space-time. Padmanabhan, in the reverse way, showed that field equations in Einstein gravity as well as in
 Lanczos-Lovelock gravity for a spherically symmetric spacetime can be expressed in the form of thermodynamic identity: $dE=TdS-PdV$. In this derivation, the modified terms could emerge in quantum pictures and hence one may think that thermodynamics can profile gravity beyond the classical level.

Alternatively, relevant to universal thermodynamics, Hayward studied thermodynamics for dynamical black hole [16,17]. He introduced the notion of
 trapping horizon in 4D Einstein gravity for non-stationary spherically symmetric spacetimes and showed that Einstein's equations are equivalent
 to the unified first law. Then projecting the unified first law along any tangential direction $ (\xi) $ to the trapping horizon, one is able
to derive the first law of thermodynamics [18-20] {\it i.e.,} Clausius relation of the dynamical black hole {\it i.e.,} $\langle A \psi,\xi \rangle = \frac{\kappa}{8 \pi G} \langle dA,\xi \rangle $, where energy flux $ \psi $ is termed as energy supply vector.

From the point of view of universal thermodynamics, we consider our universe as a non-stationary gravitational system. Further, from cosmological viewpoint the homogeneous and isotropic FRW universe may be consisdered as  dynamical spherically symmetric spacetime. Here we have only inner trapping horizon which coincides with the apparent horizon and it is possible to consider thermodynamical analysis using unified first law. Cai and Kim [21] derived the Friedmann equations with arbitrary spatial curvature starting with the fundamental relation $ \delta Q = TdS $ to the apparent horizon of the FRW universe. They have considered Hawking temperature and Bekenstein entropy respectively as
%%%%%%%%%%%%%%%%%%%%%%%%%%%%%%%%%%%%%%%%
\begin{equation}
T=\frac{1}{2\pi R_A }~,~~S=\frac{ \pi R_{A}^2 }{G}
\end{equation}
%%%%%%%%%%%%%%%%%%%%%%%%%%%%%%%%%%%%%%%%%
on the apparent horizon with $ R_A $, the radius of the apparent horizon. Further, they have shown the equivalence between the thermodynamical
laws and modified Einstein field equations in Gauss-Bonnet gravity and more general Lovelock gravity. Subsequently Cai {\it et al.} [18-20] have extensively studied unified first law in FRW universe not only for Einstein gravity but also in Lovelock gravity, Scalar-Tensor theory [18] and Brane-world scenario [20]. In this context, Eling {\it et al.} [22] have shown for $f(R)$ gravity theory that there should be entropy production term in the Clausius relation and it can be associated to shear viscosity of the horizon in pure Einstein gravity. Very recently, thermodynamical laws have been studied [23,24] in $f(R)$ gravity as well as in generalized f(R) gravity with a modified version of the entropy of the horizon. In the present work, we have modified the horizon entropy suitably so that Clausius relation is automatically satisfied.

We start with homogeneous and isotropic FRW metric as
%%%%%%%%%%%%%%%%%%%%%%%%%%%%%%%%%%%%%%%%%%
\begin{eqnarray}
ds^2&=&-dt^2 + \frac{a^2(t)}{1-kr^2}dr^2 + R^2 d\Omega_{2}^2
\nonumber
\\
&=&h_{ab}dx^a dx^b + R^2 d\Omega_{2}^2,
\end{eqnarray}
%%%%%%%%%%%%%%%%%%%%%%%%%%%%%%%%%%%%%%%%%%%
where $R=ar$ is the area radius, $h_{ab}=diag(-1,\frac{a^2}{1-kr^2})$ is the metric of 2-space $(x^0=t,x^1=r)$ and $k=0,\pm1$ denotes the curvature scalar. The above FRW metric can be written in double-null form as [18]
%%%%%%%%%%%%%%%%%%%%%%%%%%
\begin{equation}
ds^2=-2d\xi^+ d\xi^- +R^2 d\Omega_{2}^2.
\end{equation}
%%%%%%%%%%%%%%%%%%%%%%%%%%%
Here,
\begin{equation}
\partial_\pm=\frac{\partial}{\partial \xi^{\pm}}=-\sqrt{2}(\frac{\partial}{\partial t} \mp \frac{\sqrt{1-kr^2}}{a}\frac{\partial}{\partial r})
\end{equation}
%%%%%%%%%%%%%%%%%%%%%%%%%%%%%%%%%%
are future pointing null vectors. The trapping horizon (denoted by $R_T$) is defined as $\partial_+ R|_{R=R_T} =0$, which gives
%%%%%%%%%%%%%%%%%%%%%%%%%%%%%
\begin{equation}
R_T=\frac{1}{\sqrt{H^2+\frac{k}{a^2}}}=R_A.
\end{equation}
%%%%%%%%%%%%%%%%%%%%%%%%%%
The surface gravity is defined as
%%%%%%%%%%%%%%%%%%%%%%%%%%%%%%%%%%%%
\begin{equation}
\kappa=\frac{1}{2\sqrt{-h}}\partial_a(\sqrt{-h}h^{ab}\partial_b R),
\end{equation}
%%%%%%%%%%%%%%%%%%%%%%%%%%%%%%%%%%%%%%
so for any horizon (with area radius $R_h$) it can be written as
%%%%%%%%%%%%%%%
\begin{equation}
\kappa=-\left(\frac{R_h}{R_A}\right)^2\left(\frac{1-\frac{\dot{R}_A}{2HR_A}}{R_h}\right),
\end{equation}
%%%%%%%%%%%%%%%%%
and it becomes,
%%%%%%%%%%%%%%%
\begin{equation}
\kappa=-\frac{1}{R_A}\left(1-\frac{\dot{R}_A}{2HR_A}\right)=-\frac{1-\epsilon}{R_A},
\end{equation}
%%%%%%%%%%%%%%%%%%
for apparent horizon with $\epsilon=\frac{\dot{R}_A}{2HR_A}$. Note that if we assume $\epsilon<1$, then $\kappa$ is negative and hence the apparent horizon coincides with inner trapping horizon [16,18,25] (outer trapping horizon is with positive surface gravity).
The Misner-Sharp energy [16,25,26] is defined as
%%%%%%%%%%%%%%%%%%%%%%%%%%%%%%%%%
\begin{equation}
E=\frac{R}{2G}(1-h^{ab}\partial_aR\partial_bR)
\end{equation}
This is the total energy inside a sphere of radius $R$. Note that it is purely a geometric quantity and is related to the structure of the
space-time as well as to the Einstein's equations $[18]$. For the present FRW model of the universe bounded by the apparent horizon the above expression for the energy simplifies to
%%%%%%%%%%%%%%%%%%%%%%%%%%%%%%%%%%%%
\begin{equation}
E=\frac{R_A}{2G}=\frac{1}{2G\sqrt{H^2+\frac{k}{a^2}}}.
\end{equation}

{\bf f(R)-gravity}
\\

In $f(R)$ gravity, the modified Einstein-Hilbert action can be written as (in Jordan frame) [4]
%%%%%%%%%%%%%%%%%%%%%%%%%%%%%
\begin{equation}
S=\frac{1}{16\pi G} \int d^4x \sqrt{-g}f(R)+S_m,
\end{equation}
with $S_m$ as the matter action. Now, variation of $S$ with respect to the metric tensor $g_{\mu \nu}$ gives the modified field equations in $f(R)$ gravity as
%%%%%%%%%%%%%%%%%%%%
\begin{equation}
R_{\mu \nu} \frac{\partial f}{\partial R}-\frac{1}{2}g_{\mu \nu}f(R)-\nabla_\mu \nabla_\nu\left(\frac{\partial f}{\partial R}\right)+g_{\mu \nu}\nabla^2\left(\frac{\partial f}{\partial R}\right)=8\pi GT_{\mu \nu},
\end{equation}
where $T_{\mu}^\nu = diag(-\rho , p, p, p)$ is the energy-momentum tensor for the matter field in the form of perfect fluid. In particular for viable $f(R)$-gravity theory if we take
%%%%%%%%%%%%%%%%%
\begin{equation}
f(R)=R+F(R)
\end{equation}
then the explicit form of the modified field equations for FRW metric are given by
%%%%%%%%%%%%%%%%%%%
\begin{equation}
H^2+\frac{k}{a^2}=\frac{8\pi G}{3}\rho_t
\end{equation}
%%%%%%%%%%%%%%%
and
\begin{equation}
\dot{H}-\frac{k}{a^2}=-4\pi G(\rho_t+p_t),
\end{equation}
with $\rho_t=\rho +\rho_e$ and $p_t=p+p_e$.\\
The effective energy density $\rho_e$ and effective pressure $p_e$ due to the curvature contribution has the expressions
%%%%%%%%%%%%%%%%%%%%%%%%%%%%
\begin{equation}
\rho_e=\frac{1}{8\pi G}\left[-\frac{1}{2}(F-RF_1)-3H\frac{dF_1}{dt}-3F_1\left(H^2+\frac{k}{a^2}\right)\right]
\end{equation}
%%%%%%%%%%%%%%%%%%%%%%%%%%%%%%%%%%%%%%%%
\begin{equation}
\rho_e+p_e=\frac{1}{8\pi G}\left[\frac{d^2F_1}{dt^2}-H\frac{dF_1}{dt}+2F_1\left(\dot{H}-\frac{k}{a^2}\right)\right],
\end{equation}
where $R=6(\dot{H}+2H^2+\frac{k}{a^2})$ is the curvature scalar and $F_1=\frac{dF}{dR}$.\\
The energy conservation relations are
%%%%%%%%%%%%%%%%%%%%%
\begin{equation}
\dot{\rho}_m+3H(\rho_m+p_m)=0~,~~\dot{\rho}_t+3H(\rho_t+p_t)=0.
\end{equation}
So the effective pressure and energy density also satisfies the conservation relation
%%%%%%%%%%%%%%%%%%%%%%%%%%%%
\begin{equation}
\dot{\rho}_e+3H(\rho_e+p_e)=0.
\end{equation}
Following the method proposed by Cai \cite{Cai07a}, we shall derive an expression for entropy associated with the apparent horizon of a FRW universe described by the above modified Friedman equations ({\it i.e.,} Eqs. (14) and (15)). According to Refs. [16-20], the energy supply vector $\psi$ and the work density $W$ are defined as
%%%%%%%%%%%%%%%%%%%%%%%%
\begin{equation}
\psi_a=T_a^b\partial_bR+W\partial_aR~,~~W=-\frac{1}{2}T^{ab}h_{ab}.
\end{equation}
For the present model the explicit form of these quantities are
%%%%%%%%%%%%%%%%%%%%%%%%
\begin{eqnarray}
W&=&\frac{1}{2}(\rho_t-p_t)=\frac{1}{2}(\rho -p) + \frac{1}{2}(\rho_e-p_e)
\nonumber
\\
&=&W_m+W_e,
\end{eqnarray}
\begin{center}
$\psi=\psi_m+\psi_e$,
\end{center}
with
%%%%%%%%%%%%%%%%%%%%%%%%%%%
\begin{equation}
\psi_m=-\frac{1}{2}(\rho +p)HRdt+\frac{1}{2}(\rho +p)adr,
\end{equation}
%%%%%%%%%%
and
%%%%%%%%%%%%%%%%%%%%%%%%%
\begin{equation}
\psi_e=-\frac{1}{2}(\rho_e +p_e)HRdt+\frac{1}{2}(\rho_e +p_e)adr.
\end{equation}
Note that only the pure matter energy supply $A\psi_m$ (after projecting on the apparent horizon) gives the heat flow $\delta Q$ in the Clausius relation $\delta Q=TdS$, where $A=4\pi R^2$ is the surface area of a sphere of radius $R$. Thus according to Hayward [16], the (0,0) component of (modified) Einstein equations ({\it i.e.,} Eq. (14)) can be written as the unified first law
%%%%%%%%%%%%%%
\begin{equation}
dE=A\psi +WdV,
\end{equation}
where $V=\frac{4}{3}\pi R^3$ is the volume of the sphere of radius R.

 Now, using the double null vectors $\partial_\pm$ as the basis, any vector $\xi$ tangential to the apparent horizon surface can be written as
%%%%%%%%%%%%
\begin{equation}
\xi=\xi_+\partial_++\xi_-\partial_-.
\end{equation}
As by definition the trapping horizon is characterized by
\begin{center}
$\partial_+R_T=0$,
\end{center}
so on the marginal sphere,
\begin{center}
$\xi (\partial_+R_T)=0$,
\end{center}
\begin{center}
{\it i.e.,}~~ $\xi_+(\partial_+\partial_+R_T)+\xi_-(\partial_-\partial_+R_T)=0$,
\end{center}
{\it i.e.,}
%%%%%%%%%%%%%%%%%%%
\begin{equation}
\frac{\xi_+}{\xi_-}=-\frac{\partial_-\partial_+R_T}{\partial_+\partial_+R_T}.
\end{equation}
In the present model $R_T$ coincides with $R_A$ and
\begin{center}
$\partial_-\partial_+R_A=\frac{4}{R_A}(1-\epsilon)$~,~~$\partial_+\partial_+R_A=-\frac{4\epsilon }{R_A}$,
\end{center}
{\it i.e.,}
%%%%%%%%%%%%%%%%
\begin{equation}
\frac{\xi_+}{\xi_-}=\frac{1-\epsilon}{\epsilon}.
\end{equation}
Moreover using $(r,t)$ co-ordinates, $\xi$ can be written as [18]
%%%%%%%%%%%%%%%%%%
\begin{equation}
\xi =\frac{\partial}{\partial t}-(1-2\epsilon )Hr\frac{\partial}{\partial r}.
\end{equation}
Now projecting the unified first law (Eq. (24)) along $\xi$, the true first law of thermodynamics of the apparent horizon is obtained as [18,20]
%%%%%%%%%%%%
\begin{equation}
\langle dE,\xi \rangle = \frac{\kappa}{8\pi G}\langle dA,\xi \rangle +\langle WdV,\xi \rangle
\end{equation}
Note that the pure matter energy supply $A\psi_m$ when projected on the apparent horizon gives the heat flow $\delta Q$ in the Clausius relation $\delta Q=TdS$. Hence from Eq. (29) we have
%%%%%%%%%%
\begin{equation}
\delta Q=\langle A\psi_m,\xi \rangle = \frac{\kappa}{8\pi G}\langle dA, \xi \rangle -\langle A\psi_e, \xi \rangle.
\end{equation}
Using Eqs. (16), (17), (22) and (23), we obtain (after a simple algebra)
%%%%%%%%%%%%%%%
\begin{equation}
\langle A\psi_m, \xi \rangle=-\frac{2\epsilon (1-\epsilon)}{G}HR_A+\frac{A(1-\epsilon)HR_A}{8\pi G}\left(\ddot{F_1}-H\dot{F_1}+2F_1(\dot{H}-\frac{k}{a^2})\right)
\end{equation}
As the Hawking temperature on the apparent horizon is given by
%%%%%%%%%%%%%
\begin{equation}
T_A=\frac{|\kappa_A|}{2\pi}=\frac{1-\epsilon}{2\pi R_A},
\end{equation}
so the above equation can be written as
%%%%%%%%%%%%%
\begin{equation}
\langle A\psi_m, \xi \rangle =T_A\left\langle \frac{8 \pi R_A}{4G}dR_A-\frac{\pi HR_A^4}{G}\left(\ddot{F_1}-H\dot{F_1}+2F_1(\dot{H}-\frac{k}{a^2})\right)dt, \xi \right\rangle.
\end{equation}
Hence comparing with Clausius relation $\delta Q=TdS$ and integrating, we have the entropy on the apparent horizon
%%%%%%%%%%
\begin{equation}
S_A=\frac{A_A}{4G}-\frac{\pi}{G}\int \left(\ddot{F_1}-H\dot{F_1}+2F_1(\dot{H}-\frac{k}{a^2})\right)HR_A^4dt
\end{equation}
Thus the entropy on the apparent horizon differs from Bekenstein entropy by a correction term (given in the form of integral on the right hand side of the Eq. (34)).

As light rays move along the radial direction {\it i.e.,} normal to the surface of the event horizon and we have $\partial \xi^\pm=dt\mp adr$,
 one form along the normal direction, so $\partial_\pm=-\sqrt{2}(\partial_t \mp \frac{1}{a}\partial_r)$ may be chosen along the tangential direction to the surface of the event horizon. Thus for event horizon, we choose
\begin{equation}
\xi =\frac{\partial}{\partial t}-\frac{1}{a}\frac{\partial}{\partial r},
\end{equation}
as the tangential vector to the surface of the event horizon.

Now, using the expression for $\kappa$ from Eq. (7) and proceeding as before, the entropy on the event horizon turns out to be
%%%%%%%%%%%%%%%%%%%%%%%%%%%%%%
\begin{equation}
S_E=\frac{A_E}{4G}-\frac{\pi}{2G}\int \left(\frac{R_A^2 R_E}{1-\epsilon}\right)\frac{HR_E+1}{HR_E-1}\left(\ddot{F_1}-H\dot{F_1}-\frac{4F_1\epsilon}{R_A^2}\right)dR_E.
\end{equation}

Here also the leading term for entropy is the usual Bekenstein entropy.

Further, if we consider the conformal transformation
\begin{equation}
\tilde{g}_{ab}=e^{\phi}g_{ab}
\end{equation}
where the scalar field $\phi$ is defined as
\begin{equation}
 \phi \equiv lnf'(R)=ln[1+F_1 (R)],
\end{equation}
then the action $(11)$ (in Einstein frame) now becomes $[9, 27]$
\begin{equation}
 \tilde{S}=\frac{1}{16\pi G}\int [\tilde{R}-\frac{3}{2}\tilde{g}^{ab}\tilde{\nabla}_a \phi \tilde{\nabla}_b \phi -V(\phi)]\sqrt{-\tilde{g}}d^4x +S_m
\end{equation}
where $\tilde{\nabla}_a$ is the covariant derivative compatible with $\tilde{g}_{ab}$ and $V(\phi)$ is the effective potential defined as
\begin{equation}
 V=\frac{RF'(R)-F(R)}{{1+F'(R)}^2}
\end{equation}
Now varying the above action $(39)$ with respect to $\tilde{g}^{ab}$ and $\phi$ we obtain the Einstein equations (of f(R)-gravity in Einstein frame)
 and the evolution equation for $\phi$ as
\begin{equation}
 \tilde{R}_{ab}-\frac{1}{2}\tilde{g}_{ab}\tilde{R} = 3\tilde{\nabla}_a \phi \tilde{\nabla}_b \phi -\frac{1}{2}\tilde{g}_{ab}\left(\frac{3}{2}\tilde{g}^{ab}\tilde{\nabla}_a \phi \tilde{\nabla}_b \phi +V(\phi)+T_{ab}\right)
\end{equation}
and
\begin{equation}
 \tilde{\nabla}_a \tilde{\nabla}^a \phi+\frac{1}{3}\frac{\partial V}{\partial \phi}=0
\end{equation}
So for the present FRW model the explicit form of the field equations are
\begin{equation}
 H^2+\frac{k}{a^2}=\frac{8\pi G}{3}\left(\rho+ \frac{1}{8\pi G}\left(\frac{3}{4}\dot{\phi}^2+\frac{1}{2}V(\phi)\right)\right),
\end{equation}
\begin{equation}
 \dot{H}-\frac{k}{a^2}=-4\pi G\left(\rho+p+\frac{1}{8\pi G}\left(\frac{3}{2}\dot{\phi}^2\right)\right),
\end{equation}
and
\begin{equation}
 \ddot{\phi}+3H\dot{\phi}+\frac{1}{3}\frac{\partial V}{\partial \phi}=0.
\end{equation}
Thus we have
\begin{eqnarray}
 \rho_e &=&\frac{1}{8\pi G}\left(\frac{3}{4}\dot{\phi}^2+\frac{1}{2}V(\phi)\right)
\nonumber
\\
\rho_e+p_e &=& \frac{1}{8\pi G}\left(\frac{3}{2}\dot{\phi}^2\right)
\end{eqnarray}
Thus proceeding as above, the entropy on the apparent and event horizon are respectively given by
\begin{equation}
 S_A=\frac{A_A}{4G}-\frac{3\pi}{2G}\int \left(\dot{\phi}^2HR_A^4\right)dt
\end{equation}
and
\begin{equation}
 S_E=\frac{A_E}{4G}-\frac{3\pi}{4G}\int \left(\frac{R_A^2R_E}{(1-\epsilon)}\right)\left(\frac{HR_E+1}{HR_E-1}\right)\dot{\phi}^2dR_E.
\end{equation}
Note that the scalar field $\phi$ in Einstein frame corresponds to a representative form of Ricci curvature in Jordan frame. In our scenario, the
Einstein frame is the physical frame which gives self gravity of the scalar field's effective potential $V(\phi)$.\\\\

{\bf Scalar-Tensor Theory}
\\

In scalar tensor theory of gravity, using Jordan frame the Lagrangian is given by $[28]$
%%%%%%%%%%%%%%%%%%%%%%%%%
\begin{equation}
 L=\frac{1}{16\pi G}f(\phi)R-\frac{1}{2}g^{\alpha \beta}\partial_\alpha \phi \partial_\beta \phi -V(\phi)+L_m
\end{equation}
where $f(\phi)$ is an arbitrary function of the scalar field $\phi$ having potential $V(\phi)$, $L_m$ is the Lagrangian for the matter fields in the universe.\\

Now varying the action corresponding to the Lagrangian $(49)$ with respect to the dynamical variables $g_{\mu \nu}$ and $\phi$ the equation of motion are
%%%%%%%%%%%%%%%%%%%%%%%%%%%%%%%%%%%%%%%%%%%%%%%%
\begin{equation}
 G_{\alpha \beta}=\frac{8\pi G}{f(\phi)}[\partial_\alpha \phi \partial_\beta \phi -\frac{1}{2}g_{\alpha \beta}(g^{\mu \nu}\partial_\mu \phi \partial_\nu \phi)-g_{\alpha \beta}V(\phi)-g_{\alpha \beta}\nabla^2 f+\nabla_\alpha \nabla_\beta f+T_{\mu \nu}^{m}]
\end{equation}
and
%%%%%%%%%%%%%%%%%%%%%%%%%%%%%%%%%%%%%%%%
\begin{equation}
 \nabla^2 \phi -V'(\phi)+\frac{1}{2}f'(\phi)R=0
\end{equation}
%%%%%%%%%%%%%%%%%%%%%%%%%%%%%%%%%%%%%%%%%%%%%%%%%%%
where $T_{\mu \nu}^{m}$ is the energy momentum tensor of the matter distribution. Hence for FRW model, the explicit form of the equation $(50)$ and $(51)$ are
given by
%%%%%%%%%%%%%%%%%%%%%%%%%%%%%%%%%%%%%%%%%%%%%%%%%%%%%%
\begin{equation}
 H^2+\frac{k}{a^2}=\frac{8\pi G}{3f}[\rho +\frac{1}{2}\dot{\phi}^2 +V(\phi)-3Hf'\dot{\phi}]
\end{equation}
%%%%%%%%%%%%%%%%%%%%%%%%%%%
\begin{equation}
 \dot{H}-\frac{k}{a^2}=-\frac{4\pi G}{f}[(\rho +p)+\dot{\phi}^2 +(f''\dot{\phi}^2 +f'\ddot{\phi}-Hf'\dot{\phi})]
\end{equation}
%%%%%%%%%%%%%%%%%%%%%%%%%%%%%%%%%%%%%%%%%%%%%%%%
and
%%%%%%%%%%%%%%%%%%%%%%%%%%%%%%%%%%%%%%%%%%%%%%%%%
\begin{equation}
 \ddot{\phi}+3H\dot{\phi}+\frac{dV}{d\phi}=\frac{3}{8\pi G}(\dot{H}+H^2)f'
\end{equation}
%%%%%%%%%%%%%%%%%%%%%%%%%%%%%%%%%%%%%%%%%%%%%%%%%%%
Now choosing  $f(\phi)=1+F(\phi)$,the field equation $(53)$ can be rewritten as
%%%%%%%%%%%%%%%%%%%%%%%%%%%%%%%%%%%%%%%%%%%%%%
\begin{equation}
 \dot{H}-\frac{k}{a^2}=-4\pi G[(\rho +p)+\dot{\phi}^2 +(F''\dot{\phi}^2+F'\ddot{\phi}-HF'\dot{\phi})+\frac{F}{4\pi G}(\dot{H}-\frac{k}{a^2})]
\end{equation}
Hence we have
%%%%%%%%%%%%%%%%%%%%%%%%%%%%%%%%%%%%%%%%%%%%%%
\begin{equation}
 \rho_e +p_e=\dot{\phi}^2+(F''\dot{\phi}^2+F'\ddot{\phi}-HF'\phi)+\frac{F}{4\pi G}(\dot{H}-\frac{k}{a^2})
\end{equation}
%%%%%%%%%%%%%%%%%%%%%%%%%%%%%%%%%%%%%%%%%%%%%%%%%%%%%%
\\
Considering $\xi$ as given by equation $(28)$ (for apparent horizon) or by equation $(35)$ (for event horizon) and proceeding in the same way as
 before, for the validity of the unified first law, the expression of the entropy on the horizon
(apparent / event ) is given by
%%%%%%%%%%%%%%%%%%%%%%%%%%%%%%%%%%%%%%%%%%%%%%%%%
\begin{equation}
 S_A=\frac{A_A}{4G}-4\pi ^2 \int \left[ \dot{\phi}^2+\left(F''\dot{\phi}^2+F'\ddot{\phi}-HF'\phi \right)-\frac{F\epsilon}{2\pi GR_A^2}\right] \frac{R_A^3}{\epsilon}dR_A ,~
~~ \textrm{for apparent horizon}
\end{equation}
%%%%%%%%%%%%%%%%%%%%%%%%

and

%%%%%%%%%%%%%%%%%%%%%%%%%%%%%%%%%%%
\begin{equation}
 S_E=\frac{A_E}{4G}-4\pi^2 \int \left[ \left\lbrace \dot{\phi}^2+\left(F''\dot{\phi}^2+F'\ddot{\phi}-HF'\phi \right)-\frac{F\epsilon}{2\pi GR_A^2}\right\rbrace \frac{(HR_E+1)}{(HR_E-1)}\frac{R_E R_A^2}{1-\epsilon}\right]dR_E ,
~~~\textrm{for event horizon.}
\end{equation}
%%%%%%%%%%%%%%%%%%%%%%%%%%%%%%%%%%%%%%%%%%%%%%%%%%

Using similar conformal transformation as in f(R)-gravity we can write down the expressions of the entropy on the horizons in Einstein frame of scalar tensor
theory.\\\\
 {\bf Einstein Gauss-Bonnet gravity}
\\

In Einstein Gauss-Bonnet gravity, the action in (3+1) dimensions can be written as
\begin{center}
$I=\frac{1}{2}\int (\sqrt{-g}(R+\alpha G))dx^4+I_m$
\end{center}
where $\alpha$, the coupling parameter has the dimension of $(length)^2$ and $I_m$ is the matter action.
Now varying the action I over the metric tensor $g_{\mu \nu}$, we have the equations of motion: $G_{\mu \nu}-\alpha H_{\mu \nu}=T_{\mu \nu}$,
 where
\begin{center}
 $H_{\mu \nu}=4R_{\mu\lambda}R_{\nu}^\lambda+4R^{\rho \sigma}R_{\mu \rho \nu \sigma}-2RR_{\mu \nu}-2R_{\mu}^{\rho \sigma \lambda}R_{\nu \rho \sigma \lambda}+\frac{1}{2}g_{\mu \nu}G$
\end{center}
is the Lovelock tensor.
Hence for the metric given in equation $(2)$, the nonvanishing components of the modified Einstein's equations are
%%%%%%%%%%%%%%%%%
\begin{equation}
\left(H^2 +\frac{k}{a^2}\right)\left[1+\tilde{\alpha}(H^2 +\frac{k}{a^2})\right]=\frac{8\pi G \rho}{3}
\end{equation}
%%%%%%%%%%%%%%%%%%%%%%%%%%%%%%%%%%%%%%%%
and
%%%%%%%%%%%%%%%%%%
\begin{equation}
 [1+2\tilde{\alpha}(H^2 + \frac{k}{a^2})](\dot{H}-\frac{k}{a^2})=-4\pi G(\rho +p)
\end{equation}
Here $\tilde{\alpha}$ is the Gauss-Bonnet coupling parameter which is a function of $\alpha$.
Now from equation $(60)$, we have
%%%%%%%%%%%%%%%%%
\begin{equation}
 \rho_e +p_e=\frac{\tilde{\alpha}}{2\pi G}(\dot{H}-\frac{k}{a^2})(H^2+\frac{k}{a^2}),
\end{equation}

so for this modified gravity theory the expressions for the entropy on the horizon (apparent / event) is given by

%%%%%%%%%%%%%%%%%%%%%%%%%%%%%%%%%%%%%%%%%%%%%%%%%
\begin{equation}
 S_A=\frac{A_A}{4G}+\frac{4\pi}{G}\tilde{\alpha}ln(R_A)
,~~~ \textrm{for apparent horizon}
\end{equation}

and

%%%%%%%%%%%%%%%%%%%%%%%%%%%%%%%%%%%
\begin{equation}
 S_E=\frac{A_E}{4G}+\frac{4\pi \tilde{\alpha}}{G}\int \left(\frac{\epsilon}{1-\epsilon}\right)\frac{R_E}{R_A^2}\left(\frac{HR_E+1}{HR_E-1}\right)dR_E,
~~~~\textrm{for event horizon.}
\end{equation}
%%%%%%%%%%%%%%%%%%%%%%%%%%%%%%%%%%%%%%%%%%%%%%%%

Thus in the present work, we have considered universal thermodynamics for three different gravity theories (namely f(R)-gravity, Scalar-tensor theory
 and Einstein Gauss-Bonnet gravity) for FRW model of the universe bounded by apparent/event horizon. Assuming the temperature on the horizon as Hawking
temperature we have examined the validity of the unified first law and it turns out that the entropy on the horizon is no longer the Bekenstein entropy,
 rather there are correction terms in integral form. An interesting result is obtained for Einstein-Gauss-Bonnet gravity. In this modified gravity theory
the entropy of the apparent horizon achieves a logarithmic correction to Bekenstein entropy. This result is not trivial. One may get the similar
result in loop quantum gravity and also in the holographic description (one of the promising descriptions of quantum general relativity) of entropic
cosmology. Infact, for a cosmological model involving two holographic screens the universe can arrive at thermal equilibrium only after taking into
account of this logarithmic correction $[29, 30]$ . Therefore, we conclude that Universal thermodynamics in different gravity theories corresponds to
equilibrium configuration- there is no need of choosing any entropy production term, instead the entropy on the horizon is non-Bekenstein in form.

%%%%%%%%%%%%%%%%%%%%%%%%%%%%%%%%%%%%%%%%%%%%%%%%%%%%%%%%%%%%%%%%%%%%%%%%%%%%%%%%%%%%%%%%%%%%%%%%%%%%%%%%%%%%%%%%%%
\section*{Acknowledgement}
The author S.M. is thankful to UGC for NET-JRF.
The author S.S. is thankful to UGC-BSR Programme of Jadavpur University for awarding Research Fellowship.
S.C. is thankful to UGC-DRS programme, Department of Mathematics, J.U.
%%%%%%%%%%%%%%%%%%%%%%%%%%%%%%%%%%%%%%%%%%%%%%%%%%%%%%%%%%%%%%%%%%%%%%%%%%%%%%%%%%%%%%%%%%%%%%%%%%%%%%%%%%%%%%%%%%

\end{document}